\documentclass[10pt]{iopart}

\usepackage{graphicx}

\newcommand{\mathi}{{\rm i}}
\newcommand{\diff}{{\rm d}}

\renewcommand{\today}{{\bf February 28, 2006}}

\begin{document}
\title[Brueckner--Goldstone perturbation theory for 
the Hubbard model]{Brueckner--Goldstone 
perturbation theory for 
the half-filled Hubbard model in infinite dimensions}

\author{Daniel Ruhl and Florian 
Gebhard\footnote{Corresponding author: florian.gebhard@physik.uni-marburg.de}}

\address{Fachbereich Physik, Philipps-Universit\"at Marburg,
D-35032 Marburg, Germany}

\begin{abstract}
We use Brueckner--Goldstone perturbation theory to calculate
the ground-state energy of the half-filled 
Hubbard model in infinite dimensions up to fourth order 
in the Hubbard interaction. We obtain the momentum distribution
as a functional derivative of the ground-state energy with respect
to the bare dispersion relation. The resulting expressions agree with 
those from Ray\-leigh--Schr\"odinger perturbation theory.
Our results for the momentum distribution and the quasi-particle
weight agree very well with those obtained earlier from Feynman--Dyson
perturbation theory for the single-particle self-energy.
We give the correct fourth-order coefficient in the ground-state 
energy which was not calculated accurately enough from Feynman--Dyson
theory due to the insufficient accuracy of the data for
the self-energy, and find a good agreement
with recent estimates from Quantum Monte-Carlo calculations.
\end{abstract}


\submitto{JSTAT, on \today}

\maketitle

\section{Introduction}
\label{intro}

The Hubbard model~\cite{Hubbard} 
is the `standard model' for correlated electron systems~\cite{Gebhardbook}.
It contains the essential 
ingredients for the description of itinerant electrons in a solid
whose kinetic energy competes with their mutual Coulomb interaction.
Since it contains very few parameters it provides an ideal
development area and testing ground for (approximate) 
analytical and numerical techniques.

Despite its simplistic structure, the Hubbard model
is very difficult to solve exactly.
An exact solution via Bethe Ansatz exists in one spatial dimension,
see Ref.~\cite{Zebook} for a comprehensive account.
Unfortunately, an exact solution is not possible in the limit
of infinite dimension~\cite{MVPRL}, and approximate numerical and 
analytical methods must be
employed to solve the self-consistency equations of the resulting
Dynamical Mean-Field Theory~\cite{Jarrell,RMP}. 
Such approximate investigations must be supported by
perturbatively controlled analytical
calculations in the limits of weak and strong coupling.
In this way, the quality of various approximation schemes was assessed
in Refs.~\cite{alleMarburger,Eastwood,Nishimoto}.

In Ref.~\cite{alleMarburger}, Feynman--Dyson perturbation
theory was used to calculate the single-particle 
self-energy up to fourth order in the Hubbard interaction~$U$.
The ground-state energy~$E_0(U)$ and 
the momentum distribution~$n_{\sigma}(\epsilon(k);U)$ were obtained
as moments of the spectral function which cannot be expressed
in terms of a Taylor series in~$U$. Therefore, Feynman--Dyson
perturbation theory does not provide~$E_0(U)$ and~$n_{\sigma}(\epsilon(k);U)$
in form of a Taylor series in~$U$.
The coefficient of the fourth-order
term of the ground-state energy as determined in Ref.~\cite{alleMarburger} 
does not agree with results from recent Quantum Monte-Carlo 
investigations~\cite{Bluemer}. 

In this work we employ the Brueckner--Goldstone perturbation
expansion to the Hubbard model in infinite dimensions.
This (stationary) perturbation theory has the advantage
that it provides~$E_0(U)$ and~$n_{\sigma}(\epsilon(k);U)$
as a Taylor series in~$U$. As a result, the Brueckner--Goldstone approach
requires less numerical effort to calculate~$E_0(U)$ 
and~$n_{\sigma}(\epsilon(k);U)$ than the Feynman--Dyson
perturbation theory and, thus, gives more accurate results
for the same amount of computational effort.
An exception is the quasi-particle weight~$Z(U)$ which is algebraically
related to the slope of the real part of the self-energy at the Fermi energy
and gives the size of the discontinuity
of the momentum distribution at the Fermi energy. Therefore,
we can directly compare $Z(U)$ from Feynman--Dyson
perturbation theory in Ref.~\cite{alleMarburger}
with the corresponding expression from 
Brueckner--Goldstone expansion in this work.

Our work is organized as follows. In Sect.~\ref{sec:physquant},
we introduce the Hubbard model and our physical quantities
of interest. In Sect.~\ref{sec:gsener}, we use 
Brueckner--Goldstone perturbation theory to calculate the ground-state energy
of the Hubbard model in infinite dimensions up to and including
fourth order in the Hubbard interaction.
We find that the fourth-order coefficient quoted in Ref.~\cite{alleMarburger}
is indeed incorrect. The correct value agrees with recent Quantum Monte-Carlo 
data~\cite{Bluemer}.
In Sect.~\ref{sec:gsmom}, we calculate
the momentum distribution
starting from the Brueckner--Goldstone expressions for the 
ground-state energy. In infinite dimensions, 
$n_{\sigma}(\epsilon(k);U)$ can be expressed as a functional derivative
of the ground-state energy with respect to the bare dispersion 
relation~$\epsilon(k)$. 
Our results for the momentum distribution and the quasi-particle weight 
agree very well with those obtained previously from
Feynman--Dyson perturbation theory~\cite{alleMarburger}.
We conclude in Sect.~\ref{sec:discussion}.

\section{Physical quantities}
\label{sec:physquant}

\subsection{Hamilton Operator}
\label{sec:Hamilt}

We investigate spin-1/2 electrons which move 
on a $d$-dimensional lattice with $L$~lattice sites.
The corresponding operator for the kinetic energy reads
\begin{equation}
\hat{T} =  \sum_{l,m;\sigma} t(l-m) 
\hat{c}_{l,\sigma}^+\hat{c}_{m,\sigma}^{\phantom{+}} \; , 
\label{kinT}
\end{equation}
where $\hat{c}^+_{l,\sigma}$,
$\hat{c}_{l,\sigma}$ are creation and annihilation operators for
electrons with spin~$\sigma=\uparrow,\downarrow$ on site~$l$.
The thermodynamical limit, $L\to\infty$, and the limit
of infinite dimensions, $d\to\infty$, are implicitly understood henceforth.

The kinetic energy is diagonal in momentum space
\begin{equation}
\hat{T}= \sum_{k,\sigma} \epsilon(k)
\hat{c}_{k,\sigma}^+\hat{c}_{k,\sigma}^{\phantom{+}}
\end{equation}
with the dispersion relation
\begin{equation}
\epsilon(k) = \sum_{r} t(r) e^{-\mathi k r  } \; .
\end{equation}
Note that all momenta are taken from the first Brillouin zone.
The ground state of the kinetic energy is the Fermi-sea~$|{\rm FS}\rangle$,
\begin{equation}
|{\rm FS} \rangle = \prod_{k,\sigma;\epsilon(k)\leq E_{\rm F}} 
\hat{c}_{k,\sigma}^+ | {\rm vacuum}\rangle\; ,
\end{equation}
where $E_{\rm F}$ is the Fermi energy.

Later we shall work with 
a semi-circular density of states for non-interacting electrons,
\begin{equation}
D_0(\epsilon)\equiv  \frac{1}{L} \sum_k \delta(\epsilon-\epsilon(k))
= \frac{2}{\pi W}\sqrt{4 -\left(\frac{4\epsilon}{W}\right)^2\,}
\Theta\left(\left(W/2\right)^2-\epsilon^2)\right) ,
\label{rhozero}
\end{equation}
where $W\equiv 4t$ is the bandwidth and $\Theta(x)$ denotes the Heaviside 
step function.
This density of states is realized for
a Bethe lattice with infinite connectivity~\cite{Economou}.
In the following, we set $t\equiv 1$ as our energy unit.

The electron-electron interaction is taken to be 
purely on-site with strength~$U$,
\begin{equation}
\hat{H}_1 = U \hat{D} = 
U \sum_{l} \left(\hat{n}_{l,\uparrow}-\frac{1}{2}\right)
\left(\hat{n}_{l,\downarrow}-\frac{1}{2}\right) \; ,
\label{Interaction}
\end{equation}
where $\hat{n}_{l,\sigma}=
\hat{c}^+_{l,\sigma}\hat{c}_{l,\sigma}^{\phantom{+}}$ 
is the local density operator at site~$l$ for spin~$\sigma$.
The Hubbard Hamiltonian then becomes
\begin{equation}
\hat{H}= \hat{T} + \hat{H}_1 = \hat{T} + U \hat{D} \; .
\label{Hamilt}
\end{equation}
The Hamiltonian is invariant under SU(2) spin transformations
so that the total spin is a good quantum number.
We demand that the model exhibits particle-hole symmetry
on bipartite lattices,
i.e., $\hat{H}$ is assumed to be invariant under the transformation 
\begin{equation}
{\rm PH}: \qquad \hat{c}_{l,\sigma}^+ \mapsto (-1)^{l}
\hat{c}_{l,\sigma}\quad ; \quad
\hat{c}_{l,\sigma} \mapsto (-1)^{l} \hat{c}_{l,\sigma}^+ \; ,
\label{phdef}
\end{equation}
where $(-1)^l=1$ on the $A$-sites and $(-1)^l=-1$ on the $B$-sites.
Therefore, the model is altogether SO(4)-symmetric~\cite{Zebook}.
This implies that there exists a nesting vector~$Q$ such that
$\exp(\mathi Q l)=(-1)^l$ and, therefore, $\epsilon(k+Q)=-\epsilon(k)$.
We consider exclusively a half-filled band
where the number of electrons~$N=2 N_{\sigma}$ equals the
number of lattice sites~$L$ so that we may set the Fermi energy to zero,
$E_{\rm F}=0$.

\subsection{Ground-state energy, momentum distribution 
and quasi-particle weight}
\label{subsec:gsandmom}

Let $|0\rangle$ be the normalized exact ground state of~$\hat{H}$. 
As a consequence of the particle-hole symmetry~(\ref{phdef}),
the ground-state energy of the Hubbard Hamiltonian~(\ref{Hamilt})
at half band-filling is an even function of~$U$~\cite{Zebook},
\begin{equation}
E_0(U) \equiv  \langle 0 |\hat{H}|0 \rangle = E_0(-U) \; .
\end{equation}
Therefore, the Taylor expansion of~$E_0(U)$ contains even powers in~$U$ only,
\begin{equation}
E_0(U) = E_0^{(0)} + U^2 E_0^{(2)} + U^4 E_0^{(4)} + \ldots \; .
\end{equation}
For a semi-circular density of states,
the average kinetic energy of the half-filled Fermi sea is readily calculated
as
\begin{equation}
\frac{E_0^{(0)}}{L} 
\equiv 2 \int_{-2}^{0} \diff \epsilon D_0(\epsilon) \epsilon =
-\frac{8}{3\pi} \; .
\end{equation}
We shall calculate the coefficients to second and fourth order 
with the help of Brueckner--Goldstone perturbation theory
in section~\ref{sec:gsener}.

The momentum distribution is defined by
\begin{equation}
n_{k,\sigma}(U) = \langle 0 |\frac{1}{L} 
\sum_{l,m} e^{\mathi (l-m)k}
\hat{c}_{l,\sigma}^+\hat{c}_{m,\sigma}^{\phantom{+}}
|0 \rangle \; .
\end{equation}
In the limit of infinite dimensions, the momentum distribution depends
on the momentum~$k$ only through the dispersion 
relation~$\epsilon(k)$~\cite{Gebhardbook}.
Therefore, we write $n_{k,\sigma}(U) \equiv n_{\sigma}(\epsilon(k); U)
\equiv n_{\sigma}(\epsilon; U)$.
The particle-hole transformation~(\ref{phdef}) ensures that 
\begin{equation}
n_{\sigma}(\epsilon; U) = n_{\sigma}(\epsilon; -U) = 
1-n_{\sigma}(-\epsilon; U) \; .
\end{equation}
Therefore, the Taylor expansion of~$n_{\sigma}(\epsilon; U)$
contains even powers in~$U$ only,
\begin{equation}
n_{\sigma}(\epsilon; U) = n_{\sigma}^{(0)}(\epsilon)  
+ U^2 n_{\sigma}^{(2)}(\epsilon) + U^4
n_{\sigma}^{(4)}(\epsilon) + \ldots \; ,
\label{eq:momTaylor}
\end{equation}
and we may restrict ourselves to $\epsilon> 0$ in the calculation
of the momentum distribution.
The leading term in~(\ref{eq:momTaylor})
is the momentum distribution for free fermions,
\begin{equation}
n_{\sigma}^{(0)}(\epsilon)  = \Theta(4-\epsilon^2)\Theta(-\epsilon)
= \left\{ \begin{array}{lcl}
1 & \hbox{for} & -2\leq \epsilon \leq 0 \\
0 & \hbox{for} & \hphantom{-}0 < \epsilon \leq 2 
\end{array}
\right.\; . 
\label{eq:zeromomdist}
\end{equation}
We shall calculate the coefficients to second and fourth order 
with the help of Brueckner--Goldstone perturbation theory
in section~\ref{sec:gsmom}.

The discontinuity of the momentum distribution at the Fermi energy 
$E_{\rm F}=0$ is of particular interest because it gives the quasi-particle
weight~$Z(U)$. By definition,
\begin{equation}
Z(U) = n_{\sigma}(\epsilon=0^-)-n_{\sigma}(\epsilon=0^+) 
= 1- 2n_{\sigma}(\epsilon=0^+) \; ,
\end{equation}
and the Taylor expansion reads
\begin{equation}
Z(U) = 1+ Z^{(2)}\frac{U^2}{t^2} +Z^{(4)}\frac{U^4}{t^4} \ldots \; .
\end{equation}
In infinite dimensions, this quantity is related to the slope of
the self-energy at the Fermi energy via
\begin{equation}
Z(U) = \left[ 1- \left. 
\frac{\partial {\rm Re}\Sigma(\omega;U)}{
\partial \omega}\right|_{\omega=0}\right]^{-1} \; .
\end{equation}
The self-energy $\Sigma(\omega;U)$ was calculated in~\cite{alleMarburger}
to fourth order in the Hubbard interaction. Therefore, we can 
directly compare our results for $Z(U)$ from Brueckner--Goldstone perturbation
theory with those from Feynman--Dyson perturbation theory.

\section{Ground-state energy}
\label{sec:gsener}

According to Brueckner--Goldstone perturbation theory~\cite{Goldstone,Metzner},
the ground-state energy can be expressed 
as a power series in the interaction strength,
\begin{equation}
E_0-E_0^{(0)}= \langle {\rm FS} | \hat{H}_1 \sum_{j=0}^{\infty}
\left( (E_0^{(0)}-\hat{T})^{-1}\hat{H}_1 \right)^{j} 
| {\rm FS} \rangle_{\rm c}\;.
\label{goldstone-e}
\end{equation}
Here, the subscript~`c' indicates that only connected diagrams must be
included in a diagrammatic formulation of the theory. 
Note that no diagrams with Hartree bubbles appear because 
they have been subtracted explicitly in the definition of~$\hat{H}_1$.
Moreover, due to particle-hole symmetry, only odd~$j$ 
contribute to the series in~(\ref{goldstone-e}).

\begin{figure}[htbp]
\begin{center}
\includegraphics[height=3cm]{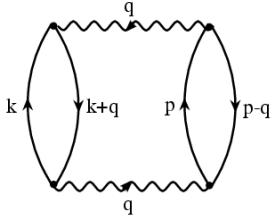}
\caption{Second-order Goldstone diagram.\label{fig:2ndorder}}
\end{center}
\end{figure}

\subsection{Second order}
\label{subsec:2ndorder-energy}

The ground-state energy to second order is given by
\begin{equation}
E_0^{(2)} = \langle {\rm FS} | \hat{D} (E_0^{(0)}-\hat{T})^{-1} \hat{D}
| {\rm FS} \rangle_{\rm c} \; ,
\end{equation}
to which only a single diagram contributes, see Fig.~\ref{fig:2ndorder}.
In this figure, a line with a down-going arrow indicates a particle line,
a line with an up-going arrow represents 
a hole line, and a wiggly line indicates the transferred momentum~$q$. 
Note that $q\neq 0$ applies because
we can write the interaction in momentum space as
\begin{equation}
\hat{D} = \frac{1}{L}\sum_{k,p}\sum_{q\neq 0} 
\hat{c}_{k+q,\uparrow}^+\hat{c}_{k,\uparrow}^{\phantom{+}} 
\hat{c}_{p-q,\downarrow}^+\hat{c}_{p,\downarrow}^{\phantom{+}} \; .
\end{equation}
The diagram is readily calculated from the Goldstone diagram rules,
\begin{eqnarray}
\frac{E_0^{(2)}}{L} &=& - \left(\frac{1}{L}\right)^3
\sum_{k,p}\sum_{q\neq 0} 
[(1-n_{\sigma}^{(0)}(\epsilon(k+q)) ]n_{\sigma}^{(0)}(\epsilon(k))
\nonumber \\
&& \hphantom{- \left(\frac{1}{L}\right)^3
\sum_{k,p}\sum_{q\neq 0} }\times
\frac{[1-n_{\sigma}^{(0)}(\epsilon(p-q))]n_{\sigma}^{(0)}(\epsilon(p))
}{\epsilon(k+q)-\epsilon(k)+\epsilon(p-q)-\epsilon(p)} \; .
\end{eqnarray}
The diagram was evaluated in one, two and three dimensions
in Ref.~\cite{Metzner}.

To make progress in the limit of infinite dimensions~\cite{MVPRL}, 
we introduce the joint density of states via
\begin{equation}
D_q(\epsilon_1,\epsilon_2) = \frac{1}{L} \sum_{k} 
\delta(\epsilon_1-\epsilon(k+q)) \delta(\epsilon_2-\epsilon(k)) \; ,
\end{equation}
and write
\begin{eqnarray}
\frac{E_0^{(2)}}{L} &=& -\!\! \int
\diff \epsilon_1\diff \epsilon_2
\diff \epsilon_3\diff \epsilon_4
\frac{[(1-n_{\sigma}^{(0)}(\epsilon_1)]n_{\sigma}^{(0)}(\epsilon_2)
[(1-n_{\sigma}^{(0)}(\epsilon_3)]n_{\sigma}^{(0)}(\epsilon_4)
}{\epsilon_1-\epsilon_2+\epsilon_3-\epsilon_4}  \nonumber \\
&& \hphantom{ - \int
\diff \epsilon_1\diff \epsilon_2
\diff \epsilon_3\diff \epsilon_4}
\frac{1}{L}
\sum_{q\neq 0} D_q(\epsilon_1,\epsilon_2)D_{-q}(\epsilon_3,\epsilon_4) \; .
\end{eqnarray}
We note that the joint density of states factorizes 
in infinite dimensions~\cite{Gebhardbook},
\begin{equation}
D_q(\epsilon_1,\epsilon_2) =  D_0(\epsilon_1) D_0(\epsilon_2)
\qquad \hbox{for `almost all'~$q$.}
\end{equation}
Therefore, the contribution of the $\uparrow$~electrons
and the $\downarrow$~electrons separate. This amounts to
the statement that the second-order diagram is a `local diagram',
i.e., the momentum conservation at the vertices
can be ignored. With the help of the Feynman trick
\begin{equation}
\frac{1}{x} = \int_{0}^{\infty}\diff \lambda e^{-\lambda x} 
\quad \hbox{for} \quad x>0
\end{equation}
we write
\begin{equation}
\frac{E_0^{(2)}}{L} = - \int_0^{\infty} \diff \lambda s_{\uparrow}(\lambda) 
s_{\downarrow}(\lambda)  \; , 
\label{eq:esecond}
\end{equation}
where
\begin{equation}
s_{\sigma}(\lambda)= F_{\rm p}(\lambda)F_{\rm h}(\lambda)
\end{equation}
with the particle and hole contributions 
\begin{eqnarray}
F_{\rm p}(\lambda)&=& \frac{1}{L} \sum_{p,\epsilon(p)>0}
e^{-\lambda\epsilon(p)} = 
\int_{0}^{\infty} \diff \epsilon D_0(\epsilon) 
e^{-\lambda\epsilon} \; ,\label{eq:Fdef}
\\
F_{\rm h}(\lambda)&=& \frac{1}{L} \sum_{p,\epsilon(p)\leq 0}
e^{\lambda\epsilon(p)} = \int_{-\infty}^{0} \diff \epsilon D_0(\epsilon) 
e^{\lambda\epsilon} \; .
\end{eqnarray}
For the calculation of the ground-state energy we may set
$F_{\rm p}(\lambda)=F_{\rm h}(\lambda)\equiv F(\lambda)$ and we arrive at
the final result~\cite{MVPRL}
\begin{equation}
\frac{E_0^{(2)}}{L} = - \int_0^{\infty} \diff \lambda [F(\lambda)]^4 \; .
\end{equation}
For the semi-circular density of states we find
the numerical value $E_0^{(2)}/L = -0.020866148383$.

\subsection{Fourth order}

\begin{figure}[htbp]
\begin{center}
\includegraphics[height=15cm]{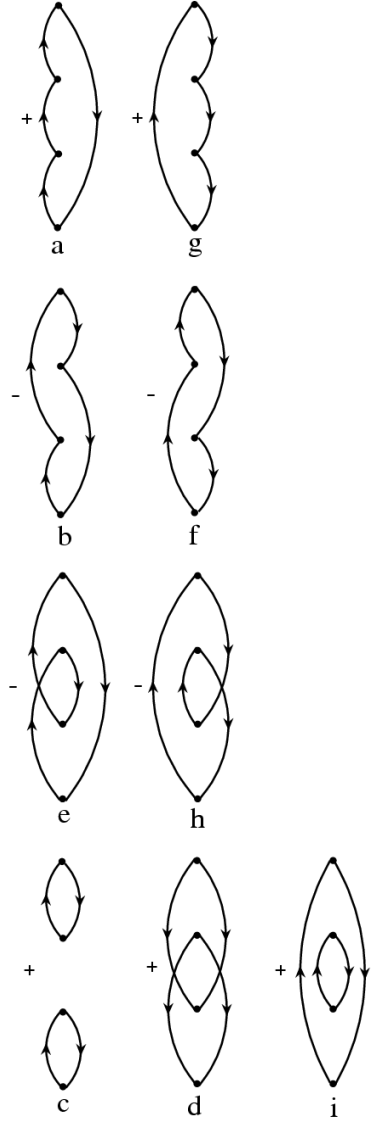}
\caption{Diagram parts for one spin species for the fourth-order 
Goldstone diagrams.\label{fig:allparts}}
\end{center}
\end{figure}

The calculation of the ground-state energy to fourth order requires
the evaluation of
\begin{equation}
E_0^{(4)} = \langle {\rm FS} | \hat{D} (E_0^{(0)}-\hat{T})^{-1} \hat{D}
(E_0^{(0)}-\hat{T})^{-1} \hat{D} (E_0^{(0)}-\hat{T})^{-1} \hat{D}
| {\rm FS} \rangle_{\rm c} \; .
\end{equation}
The Goldstone diagrams now contain four vertices and the nine
diagram parts for one spin species are shown in Fig.~\ref{fig:allparts},
together with their sign which results from the fermionic 
commutation relations after the application of Wick's theorem.
Note that the diagrams within the first, second, and third line 
are particle-hole symmetric to each other, i.e.,
$(a)\rightleftharpoons (g)$, $(b)\rightleftharpoons (f)$, 
and $(e) \rightleftharpoons (h)$.
The diagrams $(c)$, $(d)$ and~$(i)$
map onto each other under a particle-hole transformation.

\begin{figure}[htbp]
\begin{center}
\includegraphics[height=3cm]{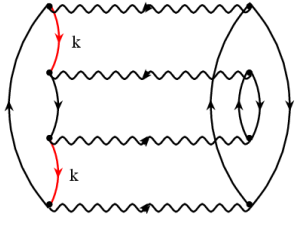}
\caption{Fourth-order Goldstone 
diagram~$(a)_{\sigma} \times (i)_{-\sigma}$
with `self-energy insertion'.\label{fig:ai-diagram}}
\end{center}
\end{figure}

In order to generate all possible diagrams, we draw all~81~possible
pairs of diagrams from~$(a)_{\uparrow} \times (a)_{\downarrow}$,
$(a)_{\uparrow} \times (b)_{\downarrow}$, \ldots,
to~$(i)_{\uparrow} \times (i)_{\downarrow}$, and connect
all vertices with horizontal interaction lines.
As an example, the diagram $(a)_{\uparrow} \times (i)_{\downarrow}$
is shown in Fig.~\ref{fig:ai-diagram}.

Not all of the 81~diagrams are connected, namely,
the diagrams $(c)_{\uparrow} \times (c)_{\downarrow}$,
$(d)_{\uparrow} \times (d)_{\downarrow}$, and
$(i)_{\uparrow} \times (i)_{\downarrow}$ are disconnected and
drop out. Moreover, the diagrams
$(a)_{\sigma}\times(c)_{-\sigma}$,
$(b)_{\sigma}\times(c)_{-\sigma}$,
$(g)_{\sigma}\times(c)_{-\sigma}$,
and
$(f)_{\sigma}\times(c)_{-\sigma}$
as well as
$(b)_{\sigma}\times(d)_{-\sigma}$
and
$(f)_{\sigma}\times(d)_{-\sigma}$
contain a hole-line and a particle-line with the same momentum
and therefore vanish due to the factor 
$n_{\sigma}^{(0)}(\epsilon)(1-n_{\sigma}^{(0)}(\epsilon))$. 
Nevertheless, it is quite
tedious to work out the remaining 66~diagrams even when
the spin-flip symmetry is employed.

We proceed differently. If all 66~diagrams were local diagrams, 
we could ignore the momentum conservation at the vertices 
as in the case of the second-order diagram in 
Sect.~\ref{subsec:2ndorder-energy}. The contribution
of all non-vanishing connected diagrams would be
\begin{eqnarray}
\frac{\widetilde{E}_0^{(4)}}{L} &=& (-1)^3 
\int_0^{\infty} \diff \lambda_1 
\int_0^{\infty} \diff \lambda_2 
\int_0^{\infty} \diff \lambda_3 \nonumber \\
&& \hphantom{(-1)^3 
\int_0^{\infty} \diff \lambda_1 }
\left[ a+b+c+d+e+f+g+h+i\right]^2
\nonumber \\
&& 
\hphantom{(-1)^3 
\int_0^{\infty} \diff \lambda_1 } - \left[c^2+d^2+e^2\right] \nonumber\\
&& \hphantom{(-1)^3 
\int_0^{\infty} \diff \lambda_1 }
- 2 \left[a+b+g+f\right]c
-2 \left[b+f\right]d\; ,
\label{eq:Wolfmodel}
\end{eqnarray}
where all functions in the integrand are functions of
$\lambda_1,\lambda_2,\lambda_3$. Explicitly, we have
\begin{eqnarray}
a(\lambda_1,\lambda_2,\lambda_3) &=& F_{\rm p}(\lambda_1+\lambda_2+\lambda_3)
F_{\rm h}(\lambda_1)F_{\rm h}(\lambda_2)F_{\rm h}(\lambda_3) \; ,\\
g(\lambda_1,\lambda_2,\lambda_3) &=& F_{\rm h}(\lambda_1+\lambda_2+\lambda_3)
F_{\rm p}(\lambda_1)F_{\rm p}(\lambda_2)F_{\rm p}(\lambda_3) \; ,\\
b(\lambda_1,\lambda_2,\lambda_3) &=& -F_{\rm p}(\lambda_1)F_{\rm h}(\lambda_3)
F_{\rm p}(\lambda_2+\lambda_3)F_{\rm h}(\lambda_1+\lambda_2) \; ,\\
f(\lambda_1,\lambda_2,\lambda_3) &=& -F_{\rm h}(\lambda_1)F_{\rm p}(\lambda_3)
F_{\rm h}(\lambda_2+\lambda_3)F_{\rm p}(\lambda_1+\lambda_2) \; ,\\
e(\lambda_1,\lambda_2,\lambda_3) &=& -F_{\rm p}(\lambda_1+\lambda_2+\lambda_3)
F_{\rm h}(\lambda_1+\lambda_2)
F_{\rm p}(\lambda_2)F_{\rm h}(\lambda_2+\lambda_3) \; ,
\nonumber \\
&& \\
h(\lambda_1,\lambda_2,\lambda_3) &=& -F_{\rm h}(\lambda_1+\lambda_2+\lambda_3)
F_{\rm p}(\lambda_1+\lambda_2)
F_{\rm h}(\lambda_2)F_{\rm p}(\lambda_2+\lambda_3) \; ,
\nonumber \\
&& \\
c(\lambda_1,\lambda_2,\lambda_3) &=& F_{\rm h}(\lambda_1)F_{\rm p}(\lambda_1)
F_{\rm h}(\lambda_3)F_{\rm p}(\lambda_3) \; ,\\
d(\lambda_1,\lambda_2,\lambda_3) &=& F_{\rm p}(\lambda_1+\lambda_2)
F_{\rm h}(\lambda_1+\lambda_2)
F_{\rm p}(\lambda_2+\lambda_3)F_{\rm h}(\lambda_2+\lambda_3) \; ,
\nonumber \\
&& \\
i(\lambda_1,\lambda_2,\lambda_3) &=& F_{\rm p}(\lambda_1+\lambda_2+\lambda_3)
F_{\rm h}(\lambda_1+\lambda_2+\lambda_3)
F_{\rm p}(\lambda_2)F_{\rm h}(\lambda_2) \; .
\nonumber \\
&& 
\end{eqnarray}
Eq.~(\ref{eq:Wolfmodel}) gives the fourth-order ground-state energy 
for the Wolff model~\cite{Wolff}
where the Hubbard interaction acts only on one site.

However, even in infinite dimensions not all diagrams are local diagrams.
First, there are diagrams with a `self-energy insertion', namely
$(a)_{\sigma}\times (i)_{-\sigma}$ as shown in Fig.~\ref{fig:ai-diagram},
and its particle-hole transformed counterpart
$(g)_{\sigma}\times (i)_{-\sigma}$.
As indicated in the figure, two hole lines have the same momentum.
Therefore, the contribution of these diagrams is
\begin{equation}
A = 2 (-1)^3 
\int_0^{\infty} \diff \lambda_1 
\int_0^{\infty} \diff \lambda_2 
\int_0^{\infty} \diff \lambda_3  \left[a_2+g_2\right] i
\end{equation}
with
\begin{eqnarray}
a_2(\lambda_1,\lambda_2,\lambda_3) &=& F_{\rm p}(\lambda_1+\lambda_2+\lambda_3)
F_{\rm h}(\lambda_1+\lambda_3)F_{\rm h}(\lambda_2)\; , \\
g_2(\lambda_1,\lambda_2,\lambda_3) &=& F_{\rm h}(\lambda_1+\lambda_2+\lambda_3)
F_{\rm p}(\lambda_1+\lambda_3)F_{\rm p}(\lambda_2) \; .
\end{eqnarray}
The contribution~$A$ replaces the terms $2(a+g)i$ in~(\ref{eq:Wolfmodel}).

\begin{figure}[htbp]
\begin{center}
\includegraphics[height=3cm]{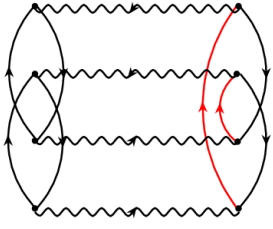}
\caption{`Pauli-forbidden' fourth-order Goldstone 
diagram~$(d)_{\sigma} \times (h)_{-\sigma}$.
The two hole-lines in the right part of the figure carry the same 
momentum.\label{fig:dh-diagram}}
\end{center}
\end{figure}

Second, different contributions result from the `Pauli-forbidden diagrams'.
The first class of Pauli-forbidden diagrams is generated by
$(d)_{\sigma} \times (h)_{-\sigma}$, as shown in Fig.~\ref{fig:dh-diagram},
and its particle-hole transformed counterpart
$(d)_{\sigma}\times (e)_{-\sigma}$.
As shown in the figure, there appear to be two holes with the same
momentum in an intermediate state.
Recall that Goldstone diagrams which appear to violate the exclusion
principle must be retained~\cite{Goldstone}.
The corresponding contribution is
\begin{equation}
B = 2 (-1)^3 
\int_0^{\infty} \diff \lambda_1 
\int_0^{\infty} \diff \lambda_2 
\int_0^{\infty} \diff \lambda_3  \left[h_2+e_2\right] d
\end{equation}
with
\begin{eqnarray}
h_2(\lambda_1,\lambda_2,\lambda_3) &=& 
-F_{\rm h}(\lambda_1+2\lambda_2+\lambda_3)
F_{\rm p}(\lambda_2+\lambda_3)F_{\rm p}(\lambda_1+\lambda_2) \; ,\\
e_2(\lambda_1,\lambda_2,\lambda_3) &=& 
-F_{\rm p}(\lambda_1+2\lambda_2+\lambda_3)
F_{\rm h}(\lambda_2+\lambda_3)F_{\rm h}(\lambda_1+\lambda_2) \; .
\end{eqnarray}
The contribution~$B$ 
replaces the terms $2(h+e)d$ in~(\ref{eq:Wolfmodel}).

\begin{figure}[htbp]
\begin{center}
\includegraphics[height=3cm]{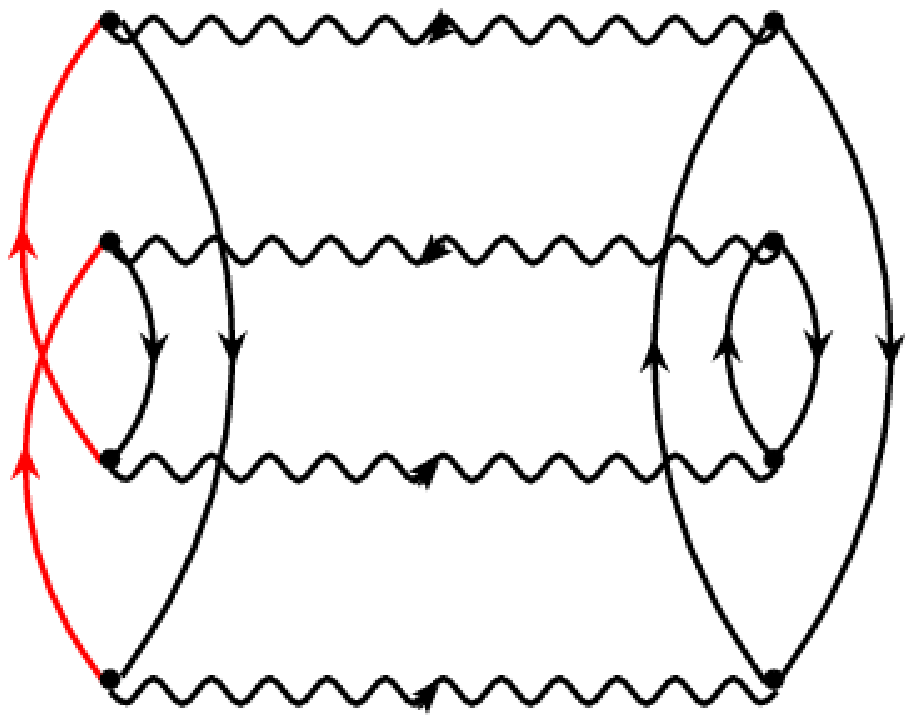}
\caption{`Pauli-forbidden' fourth-order Goldstone 
diagram~$(e)_{\sigma} \times (i)_{-\sigma}$.
The two hole lines in the left part of the figure carry the same 
momentum.\label{fig:ei-diagram}}
\end{center}
\end{figure}

The second class of Pauli-forbidden diagrams results from 
$(e)_{\sigma} \times (i)_{-\sigma}$, 
as shown in Fig.~\ref{fig:ei-diagram}, 
and its particle-hole transformed counterpart 
$(h)_{\sigma} \times (i)_{-\sigma}$,
where again two holes have the same momentum in an intermediate state.
The corresponding contribution is
\begin{equation}
C = 2 (-1)^3 
\int_0^{\infty} \diff \lambda_1 
\int_0^{\infty} \diff \lambda_2 
\int_0^{\infty} \diff \lambda_3  \left[h_4+e_4\right] i
\end{equation}
with
\begin{eqnarray}
h_4(\lambda_1,\lambda_2,\lambda_3) &=& 
-F_{\rm h}(\lambda_1+\lambda_2+\lambda_3)
F_{\rm p}(\lambda_1+2\lambda_2+\lambda_3)F_{\rm h}(\lambda_2) \; ,\\
e_4(\lambda_1,\lambda_2,\lambda_3) &=& 
-F_{\rm p}(\lambda_1+\lambda_2+\lambda_3)
F_{\rm h}(\lambda_1+2\lambda_2+\lambda_3)F_{\rm p}(\lambda_2)\; .
\end{eqnarray}
The contribution~$C$ replaces the terms $2(h+e)i$ in~(\ref{eq:Wolfmodel}).

Altogether, the ground-state energy to fourth order becomes
\begin{eqnarray}
\frac{E_0^{(4)}}{L} &=& \frac{\widetilde{E}_0^{(4)}}{L} +A+B+C \\
&& - \int_0^{\infty} \diff \lambda_1 
\int_0^{\infty} \diff \lambda_2 
\int_0^{\infty} \diff \lambda_3 \left[ 2(a+g+h+e)i+  2(h+e)d\right] \; .
\nonumber
\end{eqnarray}
The single three-fold integral can be carried out on a PC after a suitable
representation of $F(\lambda)$ in~(\ref{eq:Fdef}) has been found, e.g.,
in terms of a power series for small and large arguments.

For the semi-circular density of states, we find the fairly small value
$E_0^{(4)}/L = 7.2475 \cdot 10^{-6}$.
Thus, the ground-state energy of the Hubbard model
on a Bethe lattice with infinite connectivity reads
\begin{eqnarray}
\frac{E_0(U)}{L} &=& -\frac{8t}{3\pi} - 0.02086614838 \frac{U^2}{t}
+ 0.0000072475 \frac{U^4}{t^3} +{\cal O}\left(\frac{U^6}{t^5}\right)
\nonumber \\
                 &=& -\frac{2W}{3\pi} -0.08346459 \frac{U^2}{W}
+ 0.00046384 \frac{U^4}{W^3} +{\cal O}\left(\frac{U^6}{W^5}\right)\; ,
\nonumber \\
&&  
\label{eq:energyfinal}
\end{eqnarray}
up to fourth order in the interaction ($W=4t$ is the bandwidth).
Similar to the situation in the symmetric 
single-impurity Anderson model~\cite{YY}, the 
coefficient to fourth order is very small.
For this reason, it is very difficult to extract it from the
moments of the single-particle Green function. In fact, 
in Ref.~\cite{alleMarburger}, the single-particle
self-energy as a function of frequency was not calculated accurately enough
to give the correct sign of the fourth-order coefficient.
The correct value given in~(\ref{eq:energyfinal})
agrees very well with latest data from Quantum Monte-Carlo 
calculations~\cite{Bluemer}, $E_0^{(4),{\rm QMC}}/L = (5\pm 2) \cdot 10^{-4}$.

\section{Momentum distribution and quasi-particle weight}
\label{sec:gsmom}

As shown in Sect.~\ref{sec:gsener},
in infinite dimensions 
the ground-state energy can be cast into the form
\begin{equation}
E_0 = \sum_k\epsilon(k) n(\epsilon(k)) + \langle 0 |\hat{H}_1|0\rangle\; ,
\end{equation}
where $n(\epsilon)=2 n_{\sigma}(\epsilon(k))$ is the spin-summed
momentum distribution. This equation shows that we can obtain
$n_{\sigma}(\epsilon(k))$ as a functional derivative of the ground-state
energy,
\begin{equation}
n_{\sigma}(\epsilon(k)) = \frac{1}{2} \frac{\delta E_0}{\delta \epsilon(k)}
\; .
\label{eq:derivative}
\end{equation}
This is indeed correct to lowest order, as seen from
\begin{equation}
E_0^{(0)} = 2 \sum_{p;-2\leq \epsilon(p)\leq 0} \epsilon(p) \; ,
\end{equation}
which, together with~(\ref{eq:derivative}), 
leads to~(\ref{eq:zeromomdist}).

\subsection{Second order}

\begin{figure}[htbp]
\begin{center}
\rotatebox{-90}{\includegraphics[width=8.5cm]{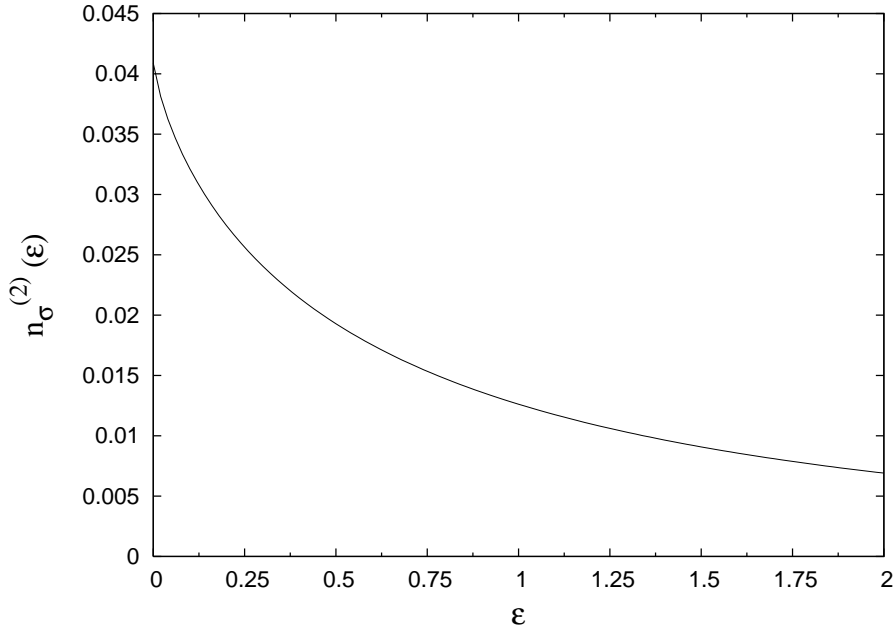}}
\caption{Second-order contribution to the momentum distribution for
$\epsilon>0$.
\label{fig:momdis2nd}}
\end{center}
\end{figure}

In the limit of infinite dimensions, the ground-state energy
can be written as a function of $F_{{\rm p},{\rm h}}(\lambda)$, 
see eq.~(\ref{eq:Fdef}). 
For $\epsilon(k)>0$, we need 
\begin{eqnarray}
\frac{\delta F_{\rm p}(\lambda)}{\delta \epsilon(k)} = -\frac{\lambda}{L} 
e^{-\lambda \epsilon(k)} \;\, , \;\, &&
\frac{\delta F_{\rm h}(\lambda)}{\delta \epsilon(k)} = 0 \;\, , \;\,
s'_{\sigma}(\lambda;\epsilon) = t(\lambda;\epsilon) F(\lambda)\; , \\
&& t(\lambda;\epsilon) \equiv \lambda e^{-\lambda \epsilon}
\; . \label{eq:abbreviation}
\end{eqnarray}
When we apply~(\ref{eq:derivative}) to the expression~(\ref{eq:esecond})
for the second-order ground-state energy
we find for the  momentum distribution to second order
\begin{equation}
n_{\sigma}^{(2)}(\epsilon) = \int_{0}^{\infty} \diff\lambda 
s_{\sigma}(\lambda) s'_{\sigma}(\lambda;\epsilon)
=  \int_{0}^{\infty} \diff\lambda\, \lambda e^{-\lambda \epsilon}
[F(\lambda)]^3 \; ,
\end{equation}
in agreement with a direct calculation from Rayleigh--Schr\"odinger
perturbation theory. We show the second-order contribution in
Fig.~\ref{fig:momdis2nd}. 

{}From the value
of $n_{\sigma}^{(2)}(\epsilon)$ at $\epsilon=0$
we deduce the second-order coefficient of the
quasi-particle weight, $Z^{(2)}=-0.0817484$.

\subsection{Fourth order}

We can readily express 
the momentum distribution to fourth order in terms of the
functions $a$ to~$i$, $a_2$, $g_2$, $h_2$, $e_2$, $h_4$, and $e_4$,
and their derivatives,
\begin{eqnarray}
n_{\sigma}^{(4)}(\epsilon) &=& \int_0^{\infty} \diff \lambda_1 
\int_0^{\infty} \diff \lambda_2 
\int_0^{\infty} \diff \lambda_3 \nonumber \\
&& \hphantom{\int_0^{\infty} \diff \lambda_1 }
\left[ a+\ldots+i\right]\left[ a'+\ldots+i'\right]- cc'-dd'-ii' 
\nonumber \\
&& \hphantom{\int_0^{\infty} \diff \lambda_1 }
- (a+b+g+f)c'- (a'+b'+g'+f')c
\nonumber \\
&& \hphantom{\int_0^{\infty} \diff \lambda_1 }
-(b'+f'+e'+h')d -(b+f+e+h)d'
\nonumber \\
&& \hphantom{\int_0^{\infty} \diff \lambda_1 }
-(a'+g'+e'+h')i -(a+g+e+h)i'
\nonumber \\
&& \hphantom{\int_0^{\infty} \diff \lambda_1 }
+ (a_2+g_2+h_4+e_4)i'+ (a'_2+g'_2+h'_4+e'_4)i
\nonumber \\
&& \hphantom{\int_0^{\infty} \diff \lambda_1 }
+ (h_2+e_2)d'+ (h'_2+e'_2)d \; . 
\label{eq:finalneps}
\end{eqnarray}
With the help of the abbreviation~(\ref{eq:abbreviation})
we can express the derivatives in the form
\begin{eqnarray}
a'(\lambda_1,\lambda_2,\lambda_3;\epsilon) 
&=& t(\lambda_1+\lambda_2+\lambda_3;\epsilon)
F(\lambda_1)F(\lambda_2)F(\lambda_3) \; ,\\
g'(\lambda_1,\lambda_2,\lambda_3;\epsilon) 
&=& F(\lambda_1+\lambda_2+\lambda_3)
\bigl[t(\lambda_1;\epsilon) F(\lambda_2)F(\lambda_3)  \nonumber \\
&&+t(\lambda_2;\epsilon) F(\lambda_1)F(\lambda_3) 
+t(\lambda_3;\epsilon) F(\lambda_1)F(\lambda_2) \bigr]\; ,\\
b'(\lambda_1,\lambda_2,\lambda_3;\epsilon) 
&=& -F(\lambda_3)F(\lambda_1+\lambda_2)
\bigl[t(\lambda_1;\epsilon)F(\lambda_2+\lambda_3) \nonumber \\
&& + t(\lambda_2+\lambda_3;\epsilon)F(\lambda_1) \bigr]\; ,\\
f'(\lambda_1,\lambda_2,\lambda_3;\epsilon) 
&=& -F(\lambda_1)F(\lambda_2+\lambda_3)
\bigl[t(\lambda_3;\epsilon)F(\lambda_1+\lambda_2)  \nonumber \\
&& + t(\lambda_1+\lambda_2;\epsilon)F(\lambda_3) \bigr]\; ,\\
e'(\lambda_1,\lambda_2,\lambda_3;\epsilon) 
&=& -F(\lambda_1+\lambda_2)F(\lambda_2+\lambda_3)
\bigl[ t(\lambda_1+\lambda_2+\lambda_3;\epsilon)F(\lambda_2) \nonumber \\
&& + t(\lambda_2;\epsilon)F(\lambda_1+\lambda_2+\lambda_3)\bigr] \; ,\\
h'(\lambda_1,\lambda_2,\lambda_3;\epsilon) 
&=& -F(\lambda_1+\lambda_2+\lambda_3)F(\lambda_2)
\bigl[t(\lambda_1+\lambda_2;\epsilon)F(\lambda_2+\lambda_3)  \nonumber \\
&& + t(\lambda_2+\lambda_3;\epsilon)F(\lambda_1+\lambda_2) \bigr] \; ,\\
c'(\lambda_1,\lambda_2,\lambda_3;\epsilon) 
&=& F(\lambda_1)F(\lambda_3)
\bigl[t(\lambda_1;\epsilon)F(\lambda_3)
+t(\lambda_3;\epsilon)F(\lambda_1)
\bigr]
 \; ,\\
d'(\lambda_1,\lambda_2,\lambda_3;\epsilon) 
&=& F(\lambda_1+\lambda_2)F(\lambda_2+\lambda_3)
\bigl[
t(\lambda_1+\lambda_2;\epsilon) F(\lambda_2+\lambda_3)  \nonumber \\
&& + t(\lambda_2+\lambda_3;\epsilon) F(\lambda_1+\lambda_2) \bigr]\; ,\\
i'(\lambda_1,\lambda_2,\lambda_3;\epsilon) 
&=& F(\lambda_1+\lambda_2+\lambda_3)F(\lambda_2)
\bigl[t(\lambda_2;\epsilon)F(\lambda_1+\lambda_2+\lambda_3) \nonumber \\
&& + t(\lambda_1+\lambda_2+\lambda_3;\epsilon)F(\lambda_2)\bigr] \; , \\
a'_2(\lambda_1,\lambda_2,\lambda_3;\epsilon) 
&=& t(\lambda_1+\lambda_2+\lambda_3;\epsilon)F(\lambda_2)F(\lambda_1+\lambda_3)
\; , \\
g'_2(\lambda_1,\lambda_2,\lambda_3;\epsilon) 
&=& F(\lambda_1+\lambda_2+\lambda_3) 
\bigl[ t(\lambda_1+\lambda_3;\epsilon)F(\lambda_2) \nonumber \\
&& +t(\lambda_2;\epsilon) F(\lambda_1+\lambda_3) \bigr] \; ,\\
h'_2(\lambda_1,\lambda_2,\lambda_3;\epsilon) 
&=& -F(\lambda_1+2\lambda_2+\lambda_3)
\bigl[t(\lambda_2+\lambda_3;\epsilon)F(\lambda_1+\lambda_2) \nonumber \\
&& + t(\lambda_1+\lambda_2;\epsilon)F(\lambda_2+\lambda_3) \bigr] \; ,\\
e'_2(\lambda_1,\lambda_2,\lambda_3;\epsilon) 
&=&-F(\lambda_2+\lambda_3)
t(\lambda_1+2\lambda_2+\lambda_3;\epsilon)F(\lambda_1+\lambda_2) \; , \\
h'_4(\lambda_1,\lambda_2,\lambda_3;\epsilon) 
&=&-F(\lambda_1+\lambda_2+\lambda_3)
t(\lambda_1+2\lambda_2+\lambda_3;\epsilon)F(\lambda_2) \; , \\
e'_4(\lambda_1,\lambda_2,\lambda_3;\epsilon) 
&=&-F(\lambda_1+2\lambda_2+\lambda_3)
\bigl[t(\lambda_1+\lambda_2+\lambda_3;\epsilon)F(\lambda_2) \nonumber \\
&& + t(\lambda_2;\epsilon)F(\lambda_1+\lambda_2+\lambda_3)\bigr] \; .
\end{eqnarray}
We note in passing that after some lengthy calculations
we obtained the same expressions from 
Rayleigh--Schr\"odinger perturbation theory~\cite{Messiah}. Thus,
stationary perturbation theory is found to be applicable
for the calculation of the momentum distribution in the thermodynamic
limit.

\begin{figure}[htbp]
\begin{center}
\rotatebox{-90}{\includegraphics[width=8.5cm]{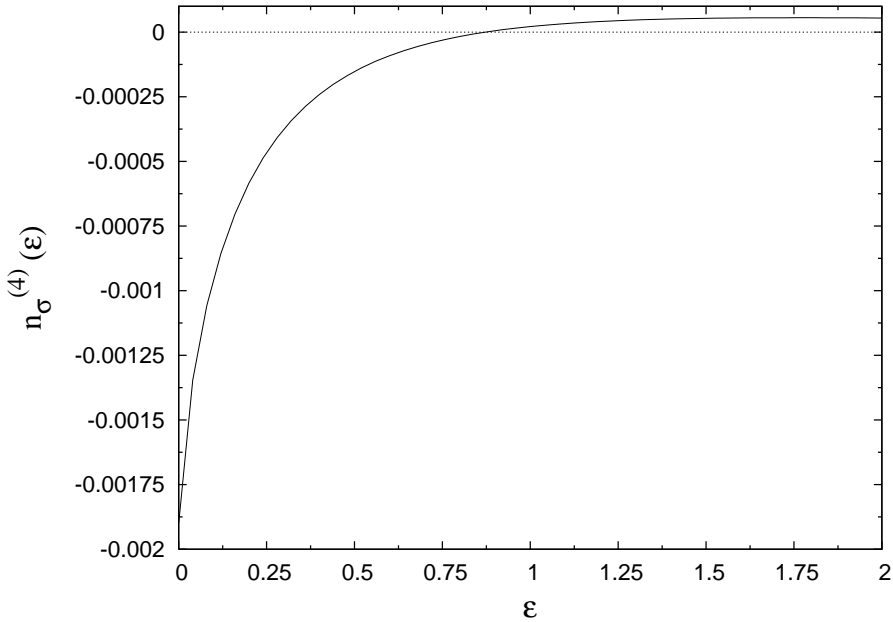}}
\caption{Second-order contribution to the momentum distribution for
$\epsilon>0$.
\label{fig:momdis4th}}
\end{center}
\end{figure}

In the actual numerical evaluation of the momentum distribution
to fourth order, it is recommendable to
separate the contributions from the local diagrams from those
of the self-energy diagram, Fig.~\ref{fig:ai-diagram}, and the two
Pauli-forbidden diagrams, Figs.~\ref{fig:dh-diagram} and~\ref{fig:ei-diagram},
which are contained in the last two lines of~(\ref{eq:finalneps}).
The latter contributions can be written as two-dimensional integrals 
because two energy denominators are identical in these
diagrams and we may use the relation
\begin{equation}
\frac{1}{x^2} = \int_{0}^{\infty}\diff \lambda \, \lambda e^{-\lambda x} 
\quad \hbox{for} \quad x>0 \; .
\end{equation}
The logarithmic singularities
for $\epsilon\to 0^+$ which arise in the three diagrams
cancel each other, and the combined integral of all terms
is convergent for all $\epsilon\geq 0$.
Note that in infinite dimensions the metallic phase obeys
the exact relation~\cite{alleMarburger}
\begin{equation}
n_{\sigma}(\epsilon\to 0^+) = \frac{1-Z(U)}{2} + \gamma(U)
\epsilon \ln(\epsilon) \; .
\end{equation}
For the first two non-trivial orders, 
the logarithmically divergent slope for $\epsilon\to 0^+$
can be seen in Figs.~\ref{fig:momdis2nd} and~\ref{fig:momdis4th}.

In Fig.~\ref{fig:momdis4th} we show the fourth-order contribution
to the momentum distribution for a semicircular density of states. 
It is seen that the correction is small
for all energies. Our results for the momentum distribution are
in very good agreement with those obtained from the single-particle
Green function~\cite{alleMarburger}. Note, however, that the
Green-function approach provides a Taylor series for the single-particle
self-energy but not for the momentum distribution.
Therefore, the numerical evaluation
of Brueckner--Goldstone perturbation theory is more
accurate than the calculation of moments of the spectral function
derived from Feynman--Dyson perturbation theory.
This applies not only for the ground-state energy but also for the
momentum distribution. 

A direct comparison of the results from
Brueckner--Goldstone and Feynman--Dyson perturbation theory
is possible for the quasi-particle weight.
As our fourth-order correction we find
$Z^{(4)}=0.00380158$. Thus, we obtain for the Hubbard model 
on a Bethe lattice with infinite connectivity
\begin{eqnarray}
Z(U) &=& 1-0.0817484 \frac{U^2}{t^2}
+ 0.00380158 \frac{U^4}{t^4} +{\cal O}\left(\frac{U^6}{t^6}\right)
\nonumber \\
                 &=& 1 -1.3079744 \frac{U^2}{W^2}
+  0.973204 \frac{U^4}{W^4} +{\cal O}\left(\frac{U^6}{W^6}\right)\; .
\label{eq:qpfinal}
\end{eqnarray}
These results compare favorably with those from Ref.~\cite{alleMarburger}
where $Z_2=1.307[1] U^2/W^2$ and $Z_4=0.969[2] U^4/W^4$ were reported.
These data also agree with recent numerical results from Quantum Monte-Carlo 
calculations~\cite{Bluemer}.

\begin{figure}[htbp]
\begin{center}
\rotatebox{-90}{\includegraphics[width=8cm]{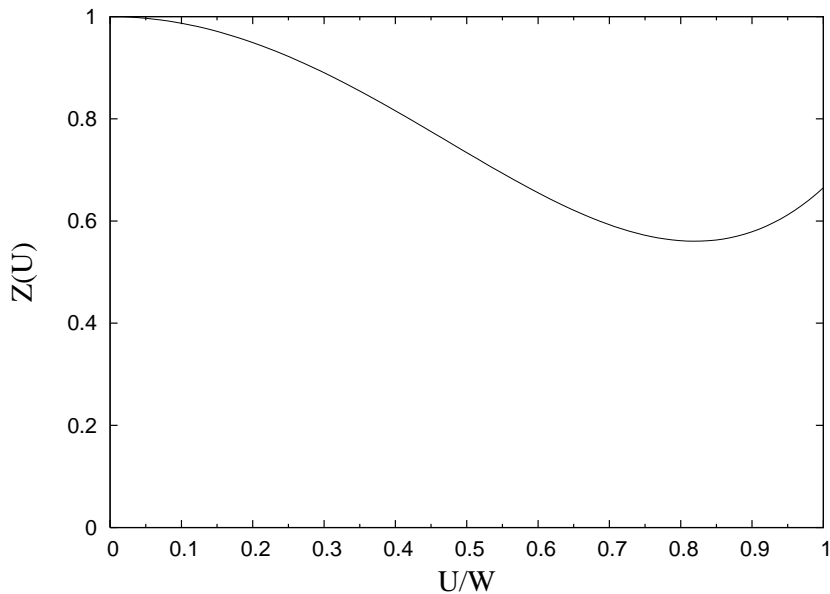}}
\caption{Quasi-particle weight as a function of the interaction strength
from~(\protect\ref{eq:qpfinal}).\label{fig:ZofU}}
\end{center}
\end{figure}

We show $Z(U)$ as a function of~$U/W$ in Fig.~\ref{fig:ZofU}.
As expected from the size of the coefficients in~(\ref{eq:qpfinal}),
the fourth-order result cannot be trusted beyond $U\approx W/2$.
This agrees with the result in Ref.~\cite{alleMarburger}. There,
the retarded self-energy up to fourth order violated the constraint
${\rm Im}\Sigma(\omega;U)\leq 0$ for $U> 0.64 W$. As expected,
fourth-order perturbation theory cannot be trusted beyond 
intermediate coupling strengths.

\section{Conclusions}
\label{sec:discussion}

In this work we employed the Brueckner--Goldstone perturbation theory 
to calculate the ground-state energy~$E_0(U)$, 
the momentum distribution~$n_{\sigma}(\epsilon(k);U)$ 
and the quasi-particle weight~$Z(U)$ 
for the Hubbard model in infinite dimensions up to and including fourth
order in the Hubbard interaction~$U$.
In infinite dimensions, $n_{\sigma}(\epsilon(k);U)$ 
can be obtained from the ground-state energy as functional derivative
with respect to the bare dispersion relation~$\epsilon(k)$. 
A straightforward but much more tedious direct calculation
via stationary Rayleigh--Schr\"odinger perturbation theory
leads to the same expressions.

Apart from the fourth-order coefficient of the ground-state energy,
we found a very good agreement of our results from Brueckner--Goldstone
perturbation theory with those obtained earlier from 
Feynman--Dyson perturbation theory~\cite{alleMarburger}.
The corrected result for the ground-state energy
agrees with recent Quantum Monte-Carlo data~\cite{Bluemer}.
In Ref.~\cite{alleMarburger}, the ground-state energy and the momentum
distribution were calculated from moments of the spectral function.
Apparently, the data for the single-particle self-energy in 
Ref.~\cite{alleMarburger} were not accurate 
enough to determine reliably the fairly small
fourth-order coefficient in the Taylor series of the ground-state energy.

Fourth-order perturbation theory becomes inapplicable beyond
$U\approx W/2$. It is prohibitive to
calculate the next order because the number of local diagrams
alone is of the order of $(200)^2$. Moreover, five-fold $\lambda$-integrations
appear in sixth order in~$U$. It should also be clear that
the sixth order would not considerably extend the region 
in~$U/W$ for which perturbation theory is applicable.
Certainly, the Mott--Hubbard transition
in infinite dimensions~\cite{Gebhardbook} cannot be addressed
perturbatively in~$U/W$.

\ack

We thank N.~Bl\"umer for providing us with his Quantum Monte-Carlo
data prior to publication.

\vspace*{6pt}

\hrule

\end{document}